\documentclass[aps,superscriptaddress,preprint,showpacs]{revtex4}
\usepackage{bm}
\usepackage{dcolumn}
\usepackage{epsfig}

\begin{document} 

\title{Meson-Meson Interactions and Resonances in
the 't Hooft Model}

\author{Zolt$\acute{{\rm a}}$n Batiz}
\author{M. T. Pe${\tilde{\rm n}}$a}
\affiliation{Centro de F\'\i sica das Intera\c c\~oes Fundamentais\\ and
Department of Physics, Instituto Superior T$\acute{e}$cnico, Av. Rovisco Pais
1049-001, Lisboa, Portugal}
\author{A.~Stadler}
  \affiliation{ 
    Centro de F\'{\i}sica Nuclear da Universidade de
    Lisboa, Av. Gama Pinto 2, P-1649-003 Lisboa, Portugal} 
  \affiliation{ 
    Departamento de F\'\i sica, Universidade de \'Evora,
    Col\'egio Lu\'\i s Verney, P-7000-671 \'Evora, Portugal}

\date{\today}

\begin{abstract}
We studied meson-meson interactions using the 't Hooft model, which represents QCD in
$1+1$ dimensions and assumes a large number of colors ($N_c$). The dominant
interactions in this large $N_c$ limit are generated by quark exchange. Our
results show that QCD in $1+1$ dimensions allows the realization of a
constituent-type quark model for the mesons, and generates a scalar ``$\sigma$''-like
meson-meson resonance, whose effective coupling and mass are determined by the
underlying QCD dynamics. These results suggest an interpretation of the lightest scalar mesons 
as  $q\bar{q} q \bar{q}$ systems.  
\end{abstract}
\pacs{13.75.Lb, 13.75.-n, 12.38.-t}

\keywords{'t Hooft model, QCD, quarks, mesons, 
resonances}

\maketitle

\section{Introduction}
\label{sec:introduction}
\noindent
At low energies, systems which are due to, or interact through, the strong
nuclear interaction may be described  by effective field theories, examples of
which are  Quantum Hadrodynamics (QHD) \cite{Walecka} and Chiral Perturbation
Theory \cite{gary}), or simple constituent quark models \cite {Isgur}. With
more or less phenomenological content, those frameworks are in general  very
successful, but there are still questions not completely answered. 

For instance, whether the  nature of the broad $\sigma$ meson corresponds
truly  to a simple quark structure, or to a resonance in the meson-meson
dynamics,  or to an unusual quark structure as a meson-glueball  combination,
or even to some combination of these, is still an open issue. Models of QCD did
not yet resolve this question quantitatively. In general for the light scalar
mesons, arguments have been advanced for the importance of a $q\bar q q \bar q$
component \cite{Close} at short distances, compatible with a dominant
meson-meson component at large distances. 
A very recent lattice calculation
\cite{Jaffe} also indicates that 
a $q \bar q q \bar q$ bound state may exist, just below threshold in
the non-exotic channel of pseudoscalar-pseudoscalar s-wave scattering.

In this work we apply the 't Hooft  model \cite{thooft}, a formulation
of QCD in 1+1 dimensions and in the large $N_c$ limit, to the meson-meson
scattering process. The 't Hooft model has no physical gluons, thus it
includes  no glueballs. Furthermore, due to the large $N_c$ limit, quark
exchange dominates gluon exchange and only a finite number of diagrams
contribute. Since the finite sum of regular contributing diagrams is unable to
produce a pole at real energies, no meson-meson bound states can be produced.

However, while meson-meson bound states are excluded in this framework, complex
energy resonant states should still be possible. It is the purpose of this
paper to calculate the meson-meson scattering amplitude based on meson-$q \bar
q$ vertex functions obtained in the 't Hooft model, and to look for low-lying
resonances. As we will show, the 't Hooft model 
indeed supports the existence of
meson-meson resonances, suggesting the relevance of the $q \bar{q}
q \bar{q}$ structure for the light scalar mesons.

A calculation of $\pi-\pi$ forward and backward scattering in the Dyson-Schwinger,
Bethe-Salpeter approach and in the rainbow-ladder approximation was presented in Ref.\
\cite{Cotanch}. It uses an effective $q\bar q$ interaction and incorporates features
of QCD. Based on this approach, a scalar meson emerges as a resonance in $\pi-\pi$
scattering. However, the calculation is performed using the Euclidian metric. In our
work, while simplifying the problem by working in 1+1 dimensions, we use the Minkowski
metric througout.

Clearly, a model in 1+1 dimensions is limited in its scope, and one has to be
very cautious when comparing its results to phenomena in the real world.
Nevertheless, for kinematic conditions of scattering processes with a
negligible component of the momentum transfer in the transverse direction, we
may conjecture that the 't Hooft model, and thus the calculation presented
here, has the main features of realistic microscopic QCD, and that its results
are valid at least qualitatively. 

Section \ref{S:form} reviews the model and introduces the calculational framework.  
Section \ref{mmscat} presents the input for the description of the meson-meson scattering
transition amplitude. Section \ref{S:results} shows the results, 
and  section \ref{S:conclusions} presents the conclusions.
 
\section{Formalism}
\label{S:form}
 
\subsection{ The `t Hooft model and the choice of gauge} 
This work  models the meson-meson interaction using the 't Hooft model. Here
we review briefly the dynamics of this model by starting
with the corresponding Lagrangian. Subsequently, we write the equations
that we solved for the one body (quark propagator) and two  body
(quark-antiquark bound state) problems.

The 
QCD Lagrangian is
\begin{equation}
{\cal L}=-\frac{1}{4}Tr \left[ G^{\mu \nu}G_{\mu \nu} \right]+ 
\bar{q}\left(i D_{\mu} \gamma^{\mu}-m_0\right)q \, ,
\label{1eq1}
\end{equation}
with the notation
\begin{eqnarray}
A^{\mu}=&&A^{\mu}_a \frac{\lambda_a}{2} \, , \nonumber\\
G_{\mu \nu}=&&\partial_{\mu} A_{\nu}-\partial_{\nu} A_{\mu}
+i\alpha \left[ A_{\mu},A_{\nu} \right] \, ,\nonumber\\ 
D_{\mu}=&&\partial_{\mu}+i\alpha A_{\mu} \, ,
\label{1eq2}
\end{eqnarray}
where $A^{\mu}_{a}$ are the gluon fields with the Lorentz index
$\mu$ and the color index $a$, the $\lambda_{a}$'s are 
the generators of the $SU(N)$ color group, $G_{\mu \nu}$ 
is the field tensor, $q$ is the 
quark field, $m_0$ is the bare quark mass and $\alpha$ is 
the quark-gluon coupling strength. Following 't Hooft \cite{thooft},
the coupling strength $\alpha$
depends on the number of colors in the following way:
\begin{equation}
\alpha=\frac{g}{\sqrt{N_c}} \, .
\label{1eq2a}
\end{equation} 

Introducing for an arbitrary two-vector $b$ the light cone variables:
\begin{eqnarray}
b_{+}=\frac{1}{\sqrt{2}}(b^0+b^1) \, ,\nonumber \\ 
b_{-}=\frac{1}{\sqrt{2}}(b^0-b^1) \, ,
\label{1eq3}
\end{eqnarray}
the scalar product of any two vectors $a$ and $b$ becomes
$a_{+}b_{-}+a_{-}b_{+}$, and the derivatives  correspond to
\begin{eqnarray}
\partial_{-}=&&\frac{\partial}{\partial x_+}=
\frac{1}{\sqrt{2}}(\partial^0-\partial^1)=
\frac{1}{\sqrt{2}}(\frac{\partial}{\partial x^0}+
\frac{\partial}{\partial x^1}) \, , \nonumber \\
\partial_{+}=&&\frac{\partial}{\partial x_+}=
\frac{1}{\sqrt{2}}(\partial^0+\partial^1)=
\frac{1}{\sqrt{2}}(\frac{\partial}{\partial x^0}-
\frac{\partial}{\partial x^1}) \, . 
\end{eqnarray}
In the same way, the $+$ and $-$ components 
of the $\gamma$ matrices are defined. The anticommutation relations read
\begin{eqnarray}
\{ \gamma_-, \gamma_- \}=\{ \gamma_+, \gamma_+ \}=0 \, , \nonumber \\
\{ \gamma_+, \gamma_- \}=2 \, .
\end{eqnarray}
In the light cone variables, the nonvanishing components 
of the field strength tensor are
\begin{equation}
G_{+-}=-G_{-+}=\partial_+A_- -\partial_
-A_+ +i\alpha[A_-,A_+] \, ,
\end{equation}
so that the free gauge field Lagrangian becomes
\begin{equation}
{\cal L}_{free}=\frac{1}{2}G_{+-}^2 \, .
\end{equation}
We choose to work in the light 
cone gauge, $A_{-}=0$, so that the commutator contained in the 
field tensor $G_{+-}$ disappears. Moreover, 
since we consider only two dimensions, 
we do not have any physical gluonic degrees of freedom. In addition, because
of the gauge condition, there is only one degree of freedom left. 
Consequently, the gluonic field is not a 
dynamical variable and does not couple to ghosts any longer.

In this parametrization the Lagrangian (\ref{1eq1}) becomes
\begin{equation}
{\cal L}=\frac{1}{2} Tr \left[(\partial_- 
A_{+})^2 \right] + 
\bar{q}\left(i \partial_+
\gamma_- +i \partial_- \gamma_+ -\alpha \gamma_{-}A_{+}  -m_0 \right)q \, .
\label{1eq4}
\end{equation}
Before quantizing the theory given by this Lagrangian, 
we calculate the gluonic field. The equation 
of motion related to this field is
\begin{equation}
(\frac{\partial}{\partial x_+})^2 A_{+}= - \alpha \bar{q} \gamma_{-} q \, .
\label{1eq5}
\end{equation}
The solution of Eq. (\ref{1eq5}) is
\begin{equation}
A_{+}(x_{+},x_{-})= - \alpha \int dy_{+} \bar{q}
(y_{+},x_{-}) \gamma_{-} q(y_{+},x_{-}) 
{\cal G}(y_{+}-x_{+}) \, ,
\label{1eq6}
\end{equation}
where the Green's function ${\cal G}$ is given by
\begin{equation}
{\cal G}(y_{+}-x_{+})=|y_{+}-x_{+}|+c_1(y_{+}-x_{+})+c_2 \, .
\label{1eq7}
\end{equation}
The coefficients $c_1$ and $c_2$ are free parameters. 
This means that the gauge 
condition did not eliminate all the redundant degrees of freedom, 
just as the Coulomb or  Lorentz gauge do not determine uniquely the photon 
propagator in QED (Gribov ambiguity). We can therefore set the 
coefficients $c_1$ and $c_2$
equal to zero
in order to simplify our calculations.

The Fourier transform of the Green's function (\ref{1eq7})
gives the gluon ``propagator'', or more precisely the 
momentum dependence of the effective quark-quark interaction:
\begin{equation}
D(k) = D(k_-)=\frac{1}{k_-^2}-\delta{(k_-)} 
{\cal P}\int_{-\infty}^{\infty} \frac{d \ell_-}{\ell^2_-} \, .
\label{1eq8}
\end{equation}

The second term in Eq. (\ref{1eq8}) was first considered
by Gross and Milana \cite{GROSS1}, in the different context
of a quasi-potential two-body equation for the quark-antiquark system. 
It makes the potential $A_+$ finite everywhere. 

From this point we proceed to solve the one body equation 
for the quark propagator.

\subsection{Quark Dyson-Schwinger equation}
\label{S:qdse}

The (undressed) 
fermion propagator is
\begin{equation}
S_0(k)=\frac{k_- \gamma_+ + k_+ \gamma_- +m_0}
{2 k_+ k_- - m_0^2 + i \epsilon} \, ,
\label{1eq9}
\end{equation}
and the quark-gluon interaction is
\begin{equation}
-i {\cal V}=-i\alpha \gamma_- \, .
\label{1eq10}
\end{equation}

\begin{figure}[hbt]
\centerline{\epsfig{file=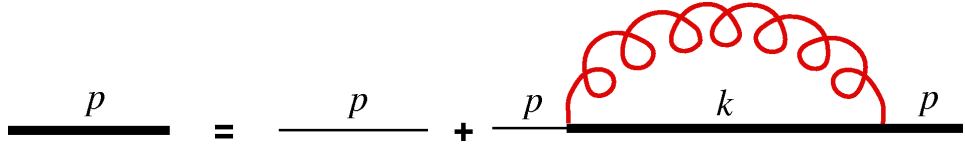,width=5.0in}}
\caption{The quark Dyson-Schwinger equation. The 
curly line represents the strong 
interaction, the thin line the unperturbed quark propagator, and 
the solid line represents the dressed quark propagator.}
\label{ds}
\end{figure}

We determine the dressed single quark propagator $S(p)$ 
using the (one body) Dyson-Schwinger equation (DSE),
\begin{eqnarray}
&&S(p)=S_0(p)+i g^2 S(p)\bigl[{\cal P} \int\,\frac{d^2k}{(2\pi)^2}\,
D(k-p)\,\gamma_- S(k)\gamma_ - \bigr]S_{0}(p) \, ,
\label{DSE}
\end{eqnarray}
which we show also graphically in Fig. \ref{ds}. 
 
Since for every internal loop there
is a factor of $\alpha^2=g^2/N_c$, and a multiplicative factor of $N_c$, the
color dependence disappears. The vertex corrections and the quark-gluon 
vertices do not have a multiplicative factor, being
supressed in the large $N_c$ limit. Therefore, in this limit the rainbow 
approximation (undressed vertices and the 
absence of the quark loops 
from the gluon propagator) is justified 
\cite{thooft}.
  
In Eq. (\ref{DSE}) $d^2k=dk_-dk_+$, and since $D(k)$ does 
not depend on $k_+$, the principal part ${\cal P} \int d^2 k $
does not depend on $p_+$. This allows  
the following parametrization of the full 
quark propagator,
\begin{equation}
S(p)=\frac{p_- \gamma_+ + \left(p_+ - 
\frac{\Sigma(p_-)}{2} \right) \gamma_-  +m_0}
{2p_-\left(p_+ - \frac{\Sigma(p_-)}{2 } \right)-m_0^2+i \epsilon} \, ,
\label{1eq11}
\end{equation}
where the self-energy contribution $\Sigma$, 
originated by ${\cal P} \int d^2 k $, depends only on
$k_-$: 

\begin{equation}
\Sigma(p_-)=-\frac{4g^2}{(2 \pi)^2 i} {\cal P}\int dk_-dk_+ D(k_- - p_-)
\frac{k_-}{2k_-k_+ -k_- \Sigma(k_-)-m_0^2+i \epsilon} \, .
\label{1eq12}
\end{equation}
Performing the $k_+$ integral first one obtains
\begin{equation}
\int dk_+ \frac{k_-}{2k_-k_+ -k_- \Sigma(k_-)-m_0^2+i \epsilon}=
-\frac{i \pi}{2} \mathrm{sgn}(k_-) \, .
\label{1eq13}
\end{equation}
Substituting this result back into Eq.\ (\ref{1eq12}) one finds that
\begin{equation}
\Sigma(p_-)=\frac{g^2}{2 \pi p_-}{\cal P} \int dk_- D(k_- - p_-)
\mathrm{sgn}(p_-) \, .
\label{1eq14}
\end{equation}
Using Eq.\ (\ref{1eq8}) for $D(k_- - p_-)$ and 
performing the integral we get           
\begin{equation}
\Sigma(p_-)=-\frac{g^2}{\pi p_-} \, .
\label{1eq15}
\end{equation}
This, in combination with Eq.\ (\ref{1eq11}) results in
\begin{equation}
S(p)=\frac{(p_+ +\frac{g^2}{2 p_- \pi} ) \gamma_- + p_- \gamma_+ + m_0 }
{2 p_+p_- - (m_0^2-\frac{g^2}{\pi} -i \epsilon)} \, .
\label{1eq16}
\end{equation}
Note that the masspole has been shifted:
\begin{equation}
m_{0}^2 \rightarrow m^2 = m_{0}^2 - \frac{g^2}{\pi}.
\label{1eq17}
\end{equation}
Having obtained the dressed propagator, 
we are ready to proceed to the next stage, namely the calculation
of the $q$-$\bar q$ bound state.

\subsection{Two-body bound states}
\label{S:2bs}

In the following, we label  
the dressed quark mass by $m_1$. As for 
the antiquark (which might have a
different flavor), we label its
dressed mass by $m_2$. 
The total momentum of the bound state is denoted by 
$r$ and the momentum of the quark by $p$.
The momentum of the antiquark 
is then $r-p$.

\begin{figure}[bt]
\centerline{\epsfig{file=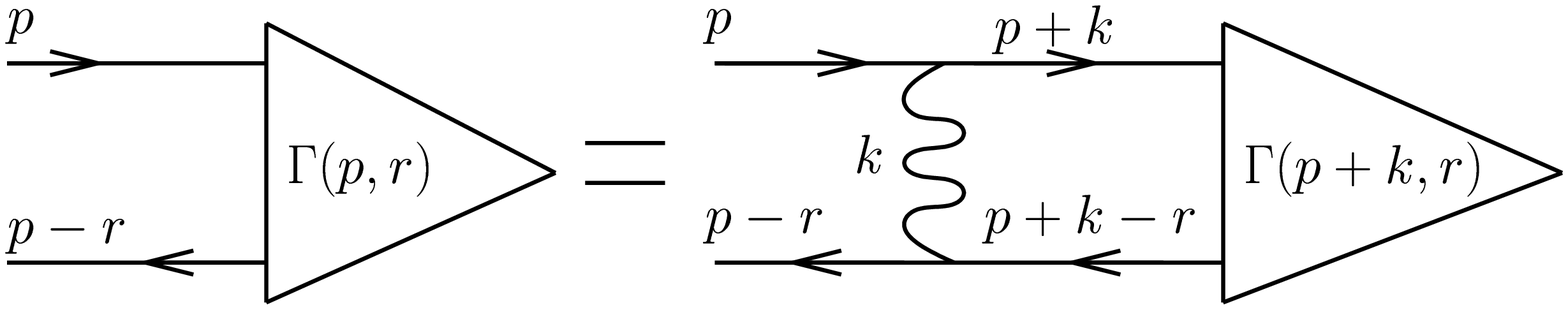,width=6.0in}}
\caption{The Bethe-Salpeter equation for the $q \bar{q}$ bound state.}
\label{bs}
\end{figure}

The bound state wave function $\Gamma(p,r)$ is given by the 
Bethe-Salpeter equation (also shown graphically in Fig. \ref{bs}),
\begin{equation}
\Gamma(p,r)=i g^2 {\cal P} \int 
\frac{d^2 k}{(2 \pi)^2}D(k_-) 
\gamma_- S_2(p+k-r) 
\Gamma(p+k,r) S_1(p+k) \gamma_- \, ,
\label{BSE}
\end{equation}
where $S_1$ and $S_2$ are the quark- 
and the antiquark propagators, respectively.
With the substitution $\Gamma(p,r)= 
\gamma_- \psi(p,r)$ \cite{thooft} Eq.\ (\ref{BSE}) becomes
\begin{equation}
\psi(p,r)=i (2g)^2 {\cal P} \int 
\frac{d^2k}{(2 \pi)^2} 
\frac{D(k_-)(p+k)_- (p+k-r)_- \psi(p+k,r)}
{\left[(p+k)^2-m_1^2 \right] \left[(p+k-r)^2-m_2^2 \right]} \, .
\label{2eq1}
\end{equation}
The equal  $x_-$ wave function is defined in the following fashion:
\begin{equation}
\varphi(p_-,r_-)=\int_{-\infty}^{\infty} dp_+ 
\frac{p_-(p-r)_-\psi(p,r)}{(p^2-m_1^2) \left[(p-r)^2-m_2^2 \right]} \, . 
\label{2eq2}
\end{equation}
By substituting this into Eq.\ (\ref{2eq1}) one gets
\begin{equation}
\psi(p,r)=\frac{g^2}{-i \pi^2} {\cal P} \int 
dk_- D(k_-) \varphi(p_-+k_-,r_-) \, .
\label{2eq3}
\end{equation}
Note that $\psi(p,r)$ does not depend on $p_+$.
Multiplying both sides of the former equation by 
$[p_-(p-r)_-]/[(p^2-m_1^2)((p-r)^2-m_2^2)]$ and integrating over $p_+$
we consider the
two poles in the complex $p_+$-plane,
namely $k_1=\frac{m_1^2}{2p_-}-i \epsilon \mathrm{sgn}(p_-)$ and 
$k_2=\frac{m_2^2}{2(p_- - r_-)}-i \epsilon \mathrm{sgn}(p_- - r_-)$. 
If both of them are in the same half plane the integral over $p_+$ is 
zero, because the sum of the two residues is zero.
If the first pole is in the upper half-plane and the second one is 
in the lower half-plane (which means that $p_- <0$ and $p_- - r_->0$)  
the integral is $2 \pi i/(k_1-k_2)=2 \pi i \theta(-p_-) 
\theta(p_- - r_-)/(k_1-k_2)$. If the second pole is in the upper half-plane
and the first one is in the lower half 
plane the integral is $-2 \pi i \theta(p_-) 
\theta(-p_- + r_-)/(k_1-k_2)$. Combining these two cases one obtains
\begin{equation}
\varphi(p,r)=\frac{\pi i}{2}
\frac{\theta(-p_-) \theta(p_- - r_-) - \theta(p_-) \theta(r_- - p_-)}
{\frac{m_1^2}{2p_-}-\frac{m_2^2}{2(p_- -r_-)}-r_+}
\psi(p,r) \, .
\label{2eq4}
\end{equation}
Whenever the combination of the $\theta$ 
functions does not vanish, it is easy to invert this relation:
\begin{equation}
\psi(p,r)=\frac{2}{\pi i}
\left(\theta(-p_-) \theta(p_- - r_-) - \theta(p_-) \theta(r_- - p_-) \right)
\left( \frac{m_1^2}{2p_-}-\frac{m_2^2}{2(p_- -r_-)}-r_+ \right)
\varphi(p,r) \, .
\label{2eq4a}
\end{equation}
Whenever this condition does not stand, we have to use 
Eq.\ (\ref{2eq3}) to compute $\psi$ from $\varphi$.
Note that
$\varphi$ has been normalized to $1/\sqrt{N_c}$ in order 
to get the correct charge normalization.

In order to determine $\varphi$, 
we transform Eq.\ (\ref{2eq2}) into a form suitable for a
numerical calculation. 
The
$\theta$ functions limit the range of $p_-$ to $0<p_-<r_-$, and
for real particles only positive values for $r_+$ must be considered.
After some algebraic manipulations,
Eq.\ (\ref{2eq4}) becomes
\begin{equation}
\mu^2 \varphi(x,r)= \left(\frac{\alpha_1}
{x}+\frac{\alpha_2}{1-x} \right) \varphi(x,r)
-{\cal P} \int_0^1 dy \frac{\varphi(y,r)-\varphi(x,r)}{(y-x)^2} \, ,
\label{2eq8}
\end{equation}
where the following notation was introduced:
\begin{equation}
\mu^2=\frac{2 \pi r_+r_-}{g^2} \, , \qquad   
\alpha_1=\frac{\pi m_1^2}{g^2} \, , \qquad   
\alpha_2=\frac{\pi m_2^2}{g^2} \, , \qquad  
x=\frac{p_-}{r_-} \, , \qquad  
y=\frac{k_-}{r_-} \, .
\label{2eq6}
\end{equation}

We solve the integral equation (\ref{2eq8}) numerically.  The wave function is
expanded in cubic splines (since the wave function $\varphi$  is defined only
in the range between $x=0$ and $x=1$, the boundary condition that they vanish
at the limits of this interval is imposed). The resulting linear matrix
equation for the expansion coefficients was solved with  standard eigenvalue
routines.

In the limit  $m_{01}=m_{02}=0$, Eq.\ (\ref{2eq8}) yields a ground state of
zero mass, and thus is consistent with chiral symmetry. To generate a solution
that is related to the pion in the real world, we searched for a bound state
solution of Eq.\ (\ref{2eq8}) with a mass of $140$ MeV. To obtain such a
solution, we varied the bare mass $m_{01}$ of one of the quarks. For
simplicity,  the second quark mass $m_{02}$ was not taken as an independent
free model  parameter but determined by assuming a fixed mass ratio
$\frac{m_{01}}{m_{02}}=3/4$, which lies within the accepted range between 
$0.2$ and $0.8$ \cite{PDG}.
The coupling parameter $g$ and the dressed masses $m_1$ and $m_2$ were adjusted accordingly,
through Eqs.(\ref{2eq8}), (\ref{2eq6}) and (\ref{1eq17}).

\begin{figure}[bt]
\vspace{5mm}
\centerline{\epsfig{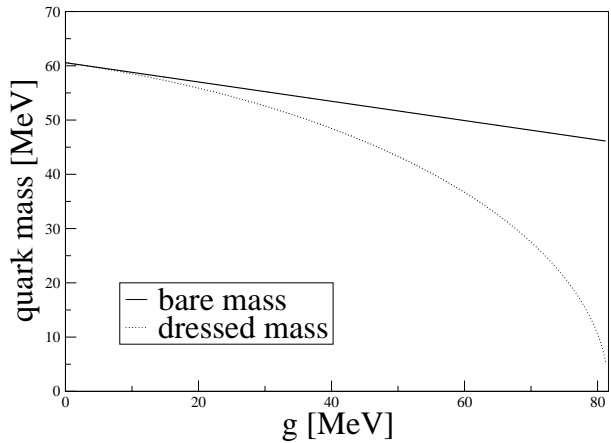}}
\caption{The bare mass (solid line) and the dressed mass (dashed line) 
of the first quark, as a function of
the strong coupling constant $g$,
with the constraints that the pion mass is $140$ MeV 
and the ratio $m_{01}/m_{02}=3/4$.}
\label{1stmass}
\end{figure}

We represent in Fig.\ \ref{1stmass} the values of the bare quark mass $m_{01}$,
as a function
of the coupling strength $g$, which generate
a bound state with a mass of $140$ MeV. 
The bare masses are found to depend linearly 
on $g$. The slope and  the $y$-axis intercept 
of the numerical straight line on Fig.
\ref{1stmass} are easily
determined through a fit, with the result
\begin{equation}
m_{01}=60.57   
-0.178 \, g \quad {\rm MeV} \, .
\end{equation} 
With the help of Eq.\ (\ref{1eq17})
we can also predict 
the dependence of the dressed masses  
on $g$ from the curves for the bare masses.
Therefore, in the 't Hooft model the dressed masses are given as the
following functions of $g$:  
\begin{eqnarray}
m_{1}^2 &=&(60.57  
-0.178  \phantom{q} g)^2-\frac{g^2}{\pi} \phantom{q}{\rm MeV}^2 \, \nonumber\\ 
m_{2}^2 &=&(80.76 
-0.24 \phantom{q} g)^2-\frac{g^2}{\pi} \phantom{q}{\rm MeV}^2 \, .
\end{eqnarray}
We can also determine the largest value of $g$, such that each dressed mass 
is physical, i.e., not imaginary. 
For the first flavor
this happens at $g=81.64$ MeV, while for the second flavor at
$g=100.8$ MeV.  The first value is therefore the largest possible for 
$g$, such the 't Hooft model may support a constituent quark model
interpretation, where the dressed masses correspond to constituent quark
masses.

\section{Meson-meson scattering}
\label{mmscat}

In this section we consider the meson-meson elastic scattering 
amplitude. We continue 
to assume two different flavors for the quarks, whose dressed masses 
are $m_1$ and $m_2$, and we consider the lowest $q\bar{q}$ bound state only.

\begin{figure}[tb]
\centerline{\epsfig{file=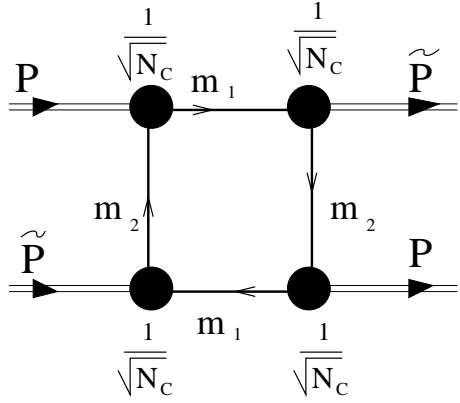,width=6cm}}
\caption{The quark-exchange box diagram. The vertex 
functions are represented by filled
circles, and the mesons by double lines.  
The power counting from the vertices is explicitely shown.
An extra factor of $N_c$ comes from the color summation in the internal loop.}
\label{box}
\end{figure}

\begin{figure}[hbt]
\centerline{\epsfig{file=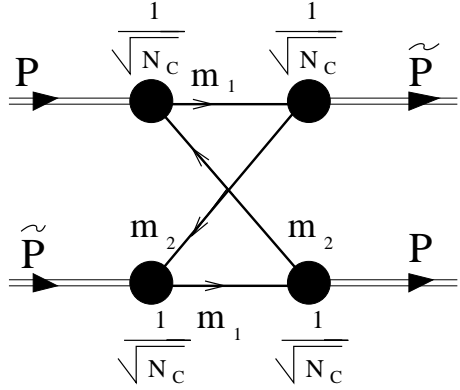,width=6cm}}
\caption{The quark-exchange crossed box diagram. As in Fig.\ \ref{box}, 
the power
counting factors are explicitely shown.
An extra factor $N_c$ comes from the color summation in the internal loop.}
\label{crossbox}
\end{figure}

The diagrams that dominate in the large $N_c$ limit are  
the quark exchange box diagram, represented in  Fig. \ref{box}, 
and the quark exchange crossbox diagram, represented in Fig. \ref{crossbox}.
In the center of mass system, the 
momenta of the ingoing mesons are $P=(P^0,P^1)=(\sqrt{\mu^2+p^2},p)$ and
${\tilde P}=({\tilde P}^0,{\tilde P}^1)=(\sqrt{\mu^2+p^2},-p)$, 
where $\mu$ is the mass of the 
meson and $p$ the relative momentum. The outgoing particles then have 
the same (but interchanged) momenta. 

Both diagrams are symmetrized in terms of the outgoing states. 
Similar diagrams which are obtained
from the former ones by interchanging $m_1$ and $m_2$ in the 
intermediate state are also
considered. There are a total of eight diagrams which were calculated. 
When there is only one quark flavor, one does not need to interchange 
the two masses and there are only four diagrams.
The sum of these 
diagrams is proportional to $1/N_c$.

\begin{figure}[tb]
\centerline{\epsfig{file=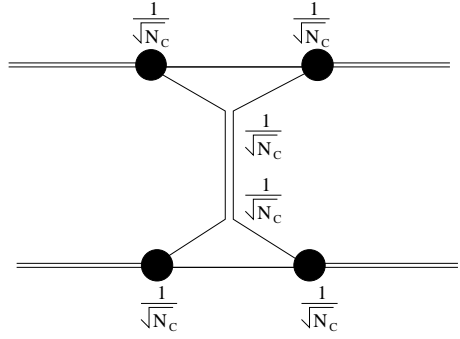,width=6cm}}
\caption{The gluon exchange diagram. Power
counting is shown as previously. For this purpose the
gluon line is represented as two parallel quark lines.
An extra factor $N_c$ comes from the color summation in the internal loop.}
\label{gluonexchange}
\end{figure}

As for the gluon exchange diagrams, such as in Fig. \ref{gluonexchange},
they are suppressed in the large $N_c$ limit by a 
factor of $1/N_c$ compared to the
quark exchange terms.

Since the vertex function is independent of the 
$+$ component of the relative momentum, 
the momentum integral in the loops of Figs. \ref{box} and \ref{crossbox}
is simplified: we can first integrate the propagator products over
$k_+$ analytically, and then perform numerically the second integration over
$k_-$, which includes now the vertex functions.

As an illustrative example,  we demonstrate the calculation of the box diagram
(Fig. \ref{box}) in greater detail.

The corresponding scattering amplitude is
\begin{eqnarray}
{\cal M}_\mathrm{box}&=&\int_{\infty}^{\infty} d k_- 
\int_{\infty}^{\infty} d k_+ 
\psi(-k,-P) 
\psi(k,{\tilde P}) \psi(P + {\tilde P}-k,P) 
\psi(k-P-{\tilde P},-{\tilde P}) \nonumber\\ 
& \times & \frac{1}{k_+-\frac{m_1^2}{2k_-}+i \epsilon \, \mathrm{sgn}(k_-)}
\frac{1}{k_+ - {\tilde P}_- - \frac{m_2^2}{2(k_- - {\tilde P}_-)}+i 
\epsilon \, \mathrm{sgn}(k_- - {\tilde P}_-)} \nonumber\\ 
& \times & \frac{1}{k_+ - P_+ - {\tilde P}_+ - \frac{m_1^2}
{2(k_- - P_- - {\tilde P}_-)}+
i \epsilon \, \mathrm{sgn}(k_- - P_- - {\tilde P}_-)} \nonumber\\ 
& \times & \frac{1}{k_+ - P_+ - \frac{m_2^2}{2(k_- - P_-)}
+i \epsilon \, \mathrm{sgn}(k_- - P_-)} \, . 
\label{4eq1}
\end{eqnarray}
The propagators have four poles:
\begin{eqnarray}
k_1 &=& \frac{m_1^2}{2k_-}-i \epsilon \, \mathrm{sgn}(k_-) \, ,\nonumber\\ 
k_2 &=& {\tilde P}_+ + \frac{m_2^2}{2(k_- - {\tilde P}_-)} -i 
\epsilon \, \mathrm{sgn}(k_- - {\tilde P}_-) \, ,\nonumber\\ 
k_3 &=& P_+ + {\tilde P}_+ + \frac{m_1^2}{2(k_- - P_- - {\tilde P}_-)}-
i \epsilon \, \mathrm{sgn}(k_- - P_- - {\tilde P}_-) \, ,\nonumber\\ 
k_4 &=& P_++\frac{m_2^2}{2(k_- - P_-)}-i \epsilon \, 
\mathrm{sgn}(k_- - P_-) \, .
\label{4eq2}
\end{eqnarray}
In order to perform the $k_+$ integration, one needs 
to close the contour in the complex plane and consider the residues of all
poles inside the contour. 
There are $16$ different possible combinations 
of signs of the imaginary parts of the poles.
Some of these cases can be excluded, because they correspond to
values of $k_-$ which make the $k_+$ integral vanish.

For instance, a
pole $k_1$ in the upper half plane implies that the  pole $k_2$
cannot be in the lower half-plane, 
otherwise one would have $k_- > \tilde{P}_-=
(\sqrt{\mu^2+p^2}+p)/\sqrt{2}>0$, in contradiction with the initial
hypothesis $k_{-}<0$.
Likewise the poles $k_3$ and $k_4$ cannot be in the lower half-plane
either. Therefore, if $k_1$  is in the upper half-plane, 
the other 3 poles are also in the  upper half plane. This would imply the
$k_+$ integral to vanish, since one may close the contour 
below the $k_+$ axis. Therefore we can exclude 
the case when $k_1$ is in the upper half plane.

After a detailed analysis one finds that there 
are only three cases that have a non-vanishing contribution to the integral: 
(i) only $k_3$ is in the upper half plane,
(ii) the poles $k_2$ and $k_3$ are in the upper half plane,
(iii) only $k_1$ is in the lower half plane.
As for case (i), it implies $k_- > {\tilde P}_-$ 
and $k_-< P_- + {\tilde P}_-$. 
Under these
circumstances, $\psi(P+{\tilde P}-k,P)=\frac{2}{\pi i} (k_3-k_2)|_{\epsilon=0} 
\varphi((P+{\tilde P}-k)_-/P_-)$ and $\psi(k-P-{\tilde P},-{\tilde P})
=\frac{2}{\pi i} 
(k_2-k_1)|_{\epsilon=0} \varphi((P+{\tilde P}-k)_-/{{\tilde P}_-})$,  
due to Eq.~ (\ref{2eq4a}), 
while the other vertex functions have to be 
evaluated using Eq.~(\ref{2eq3}). 
The contribution from case (i) becomes
\begin{eqnarray}
{\cal M}_\mathrm{box}|_1 &=& -2 \pi i \left( \frac{2}{\pi} \right)^2 
\int_{{\tilde P}_-}^{{\tilde P}_- + P_-} d k_- 
\frac{1}{(k_3-k_4)|_{\epsilon=0}} \nonumber\\ 
\times && \varphi(\frac{(P+{\tilde P}-k)_-}{{P_-}}) 
\varphi(\frac{(P+{\tilde P}-k)_-}
{{\tilde P}_-}) \psi(-k,-P) \psi(k,{\tilde P}) \, .
\label{eq36}
\end{eqnarray}
This integral is computed numerically. We treat the other two cases
in a similar fashion.

We note that the amplitude of Eq.\ (\ref{eq36})
vanishes in the chiral limit, which supports
that the ground state
solution for the $q {\bar q}$ system has features of the pion.
Indeed, since in the chiral limit Eq.\ (\ref{2eq8}) gives 
$\mu=0$, one has $P_-=\frac{1}{\sqrt{2}}(\sqrt{\mu^2+p^2}-p)=0$, 
implying that the two integration limits in (\ref{eq36}) coincide, 
and therefore 
the amplitude vanishes. The same  
can be shown for the other terms of the amplitude not explicitly written
here.

It is worth mentioning that we implemented stability tests
of the numerical results against the number of gridpoints, 
the number of splines and the singularity regulator $\epsilon$ 
(the $k_+$ integral 
of the propagators is singular). These checks proceeded by 
imposing the following 
criteria: doubling each of the mentioned parameters, the relative 
change in the results should be less than 1\% .  
Convergence is typically obtained for 440 
gridpoints, 40 splines and $\epsilon= 10^{-2}$.

\section{Results}
\label{S:results}

The numerical calculation of the meson-meson scattering  amplitude (section
\ref{mmscat}) starts with the evaluation of the two-body quark-antiquark
wavefunction from the  Bethe-Salpeter equation (section \ref{S:2bs}).  
In turn, this
demands as input the bare quark masses and the quark-gluon coupling constant
$g$ (section \ref{S:qdse}).  

We constructed four representative models which correspond to
four different choices of the quark-gluon coupling 
constant and bare quark masses.
They are subjected to the constraint that the bound state mass of the
$q\bar q$ system is the pion mass $m_{\pi}$=140 MeV.

\begin{center}
\begin{table}[bt]
\begin{ruledtabular}
\begin{tabular}{lcccc}
 & Model I & Model II & Model III & Model IV \\
 \hline
 $g$  & 1 & 20.1 & 80 & 2500 \\
 $m_{01}$ & 60.0 & 57.0    & 46.4   & 6.0 \\
 $m_{02}$ & 80.0 & 76.1    & 61.8   & 5.0 \\
 $m_1$    & 60.0 & 55.9 & 10.6 & - \\
 $m_2$    & 80.0 & 75.2 & 42.2 & - \\
\hline
$m_r$ ($=E_R$) & 280.0 & 282.4 & 280.7 & 300.0 \\
$m_i$  & 4.4 & 68.6 & 80.7 & 139.3 \\
$\Gamma$  & 0.07 & 16.4 & 23.2 & 64.7 \\
\end{tabular}
\end{ruledtabular}
\caption{Constants and resonance parameters for models I--IV. 
The first five lines show the
quark-gluon coupling constant, the bare the 
dressed quark masses (the latter are unphysical in model
IV and thus not included). The following 
lines are the results of the s-channel resonance fit using Eq.\ (\ref{3eq1}). 
All parameters are in MeV.} 
\label{models} 
\end{table}
\end{center}

The parameters defining the four models are shown in Table \ref{models}. 
The organization
principle for these models is that, going from  model I 
to model IV, the quark-gluon
coupling constant increases and the quark masses decrease. 

In models I and II,  the sum of the dressed quark masses is
close to the real pion mass. Consequently, they can be interpreted as
constituent quark masses in a constituent quark model for the ``pion'',
generated by  QCD within the 't Hooft model assumptions.
However, this correspondence breaks down for large couplings,
as seen for the model considered next.

In model III, the value of the coupling $g$ is chosen slightly 
below the maximum value
determined in section 2 for a constituent quark 
model interpretation, while in model IV
that maximum value is exceeded substantially. In 
this last case, the dressed masses are
imaginary and thus not physical.

Not surprisingly, the bare quark masses differ from the up and 
down current quark masses
of QCD for all models. This is a known artifact of the 't 
Hooft model:  from Eq.\
(\ref{1eq17}) one can see that the mass shift due to the 
dressing decreases the quark mass, instead of
increasing it as in 3+1 dimensional QCD. 

\begin{figure}[tb]
\centerline{\epsfig{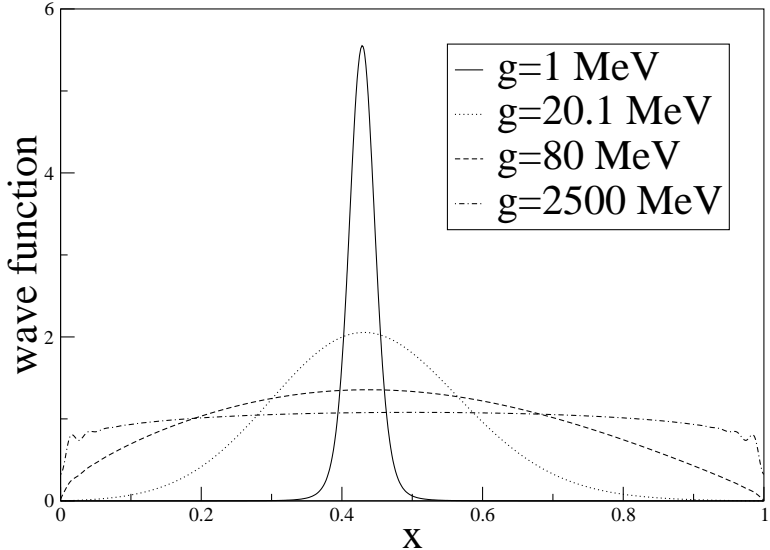}}
\caption{Bethe-Salpeter $q \bar{q}$ wave functions for models I--IV. 
$x=\frac{p_-}{r_-}$ is the momentum fraction (or Bjorken variable) 
defined in Eq.\ (\ref{2eq6}).}
\label{wavefunctions}
\end{figure}

Figure \ref{wavefunctions} shows the obtained Bethe-Salpeter $q\bar{q}$ 
wave functions  for
each model. Clearly, the description of the quark-gluon vertices 
varies considerably. The
wave functions are strongly peaked around $x$ close to 0.4 for 
small values of $g$, while
for larger $g$ they become broader and more and more constant. 
This behavior can be easily understood:
larger values of $g$ cause stronger attraction between quark 
and antiquark, leading to a 
tighter bound state and therefore a more spread out wave 
function in momentum space. 

\begin{figure}[tb]
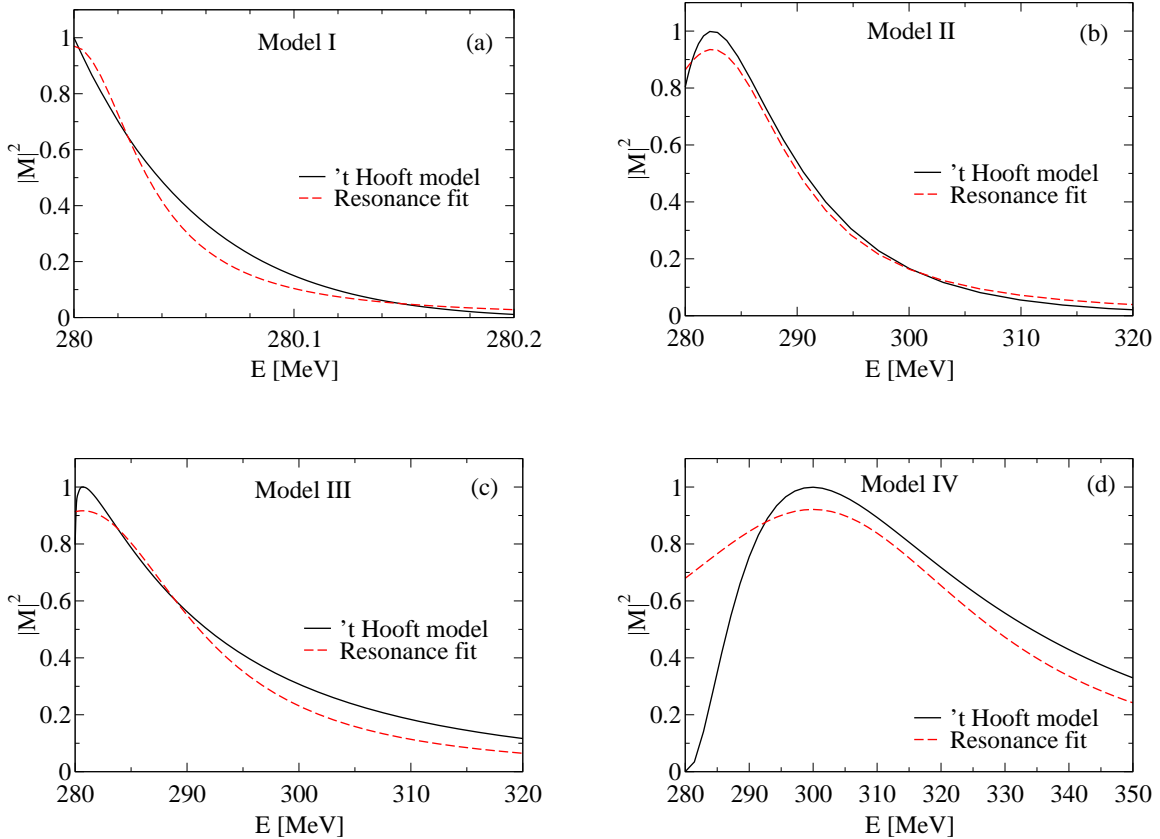
 \begin{center} $\begin{array}{c@{\hspace{1cm}}c}
\epsfig{file=AS-model1-2.eps,width=2.80in} &
\epsfig{file=AS-model2-2.eps,width=2.8in} \\ 
\\ \epsfig{file=AS-model3-2.eps,width=2.8in} &
\epsfig{file=AS-model4-2.eps,width=2.8in} \\ 
\end{array}$ \caption{Absolute squares of
meson-meson scattering  amplitudes for models I -- IV  close to threshold
energies. A resonance structure is clearly visible. The solid lines  are
obtained using the 't Hooft model, the dashed lines represent fits to a
simple resonance model.  The amplitudes are scaled  to 1 at their maximum.}
\label{F:results} \end{center} \end{figure}

As described in the previous section, for each of these  models we
calculated the meson-meson scattering amplitude, the squares of  which are
displayed in Fig.\ \ref{F:results}.  Since we are primarily interested in
their structure, the amplitudes have been scaled such that the maximum of
their absolute  squares are equal to 1. In all four cases, we find a
resonance structure close to threshold. 

This feature could be a sign for the existence of a $q^2 \bar q^2$ bound
state, for which the lattice calculations of Ref.\ \cite{Jaffe} found
indications. We remind the reader that working in perturbation theory we
cannot generate a bound state directly, but it would be interesting to see
if in a non-perturbative extension of our calculation, such a bound state
would also emerge from the 't Hooft model.

On the other hand, the experimentally observed resonances have energies well
above threshold. It may be necessary to include gluon exchange (not
considered in our calculation)  and higher order quark exchange terms
in the expansion in powers of $1/N_c$ in order to achieve a description
resembling more closely the real world.

In order to determine approximately the position and width 
of the resonance, we compare
the 't Hooft model amplitudes to a simple resonance model. 
We calculate the amplitude
for an intermediate s-channel resonance at tree level, 
the absolute square of which is
\begin{equation}
|{\cal M}|^2={\tilde g}^4\frac{1}{(s-m_r^2)^2+m_i^4} \, ,
\label{3eq1}
\end{equation}
where $m_r^2$ and $m_i^2$ are the real part 
and the imaginary part of the
square of the resonance mass, respectively,
and ${\tilde g}$ is the effective meson-meson-``$\sigma$'' coupling constant.
We then fit the parameters of this simple model to best 
reproduce the 't Hooft model
results.

In the non-relativistic limit, (\ref{3eq1}) becomes  
\begin{equation}
|{\cal M}|^2 \rightarrow {\tilde g}^4\frac{\frac{1}{4E_R^2}}{(E-E_R)^2+
\frac{m_i^4}{4E_R^2}} \, .
\label{3eq1a}
\end{equation}
We can compare (\ref{3eq1a}) with the well-known Breit-Wigner form,  
\begin{equation} 
|{\cal M}|^2 \approx {\bar
g}^4\frac{\Gamma^2/4}{(E-E_R)^2+\Gamma^2/4} \, , 
\label{3eq2} 
\end{equation}
where $E_R$ is the resonance energy, $\Gamma$ 
is its width, and ${\bar g}$ the corresponding coupling strength, 
and simply read off the relations $\Gamma^2=m_i^4/m_r^2$ and $E_R=m_r$.

The parameters obtained in this way for all models are also  displayed in Table
\ref{models}. We verified in an independent fit directly to the Breit-Wigner 
form (\ref{3eq2}) that the parameters are not significantly altered in the
non-relativistic limit. For comparison, the mass of  the ``real'' sigma
resonance is  considered to be in the range of 400--1200 MeV,  while its width
lies in the  interval 600--1000 MeV \cite{PDG}.  Clearly, one should not demand
too much from the  't Hooft model with its simplifying assumptions. However, we
consider it a significant  finding that it predicts a low energy resonance at
all, based solely on the leading-order quark exchange diagrams.

In all four cases, the resonance is located very close to threshold, and it is
very narrow.  The width increases slowly with increasing  quark-gluon coupling
strength. On the other hand, the resonance position remains more or less
unchanged as long as we restrict ourselves to models with real dressed masses 
(models I to III). Only for model IV, whose coupling constant is considerably
larger, the resonance moves away from threshold to about 300 MeV. 

One might have expected the resonance energy to increase smoothly and in a more pronounced
manner with the quark-gluon coupling strength $g$. However, one has to keep in mind that the
included quark-exchange processes do not directly depend on $g$, but only indirectly through
changes of the vertex functions and of the dressed quark masses that appear in the quark
propagators. Owing to the already mentioned peculiar feature of the 't Hooft model that
dressed quark masses decrease with increasing coupling strength $g$, the latter tend to
decrease the resonance energy. Moreover, while a larger $g$ implies an effectively stronger
attraction between the two mesons -- once they overlap -- through the stronger quark-quark
attraction, the very probability of this overlap drops in turn, because the spatial size of
the mesons decreases. These effects seems to counterbalance each other to a large extent,
leaving the resonance position more or less unchanged.

Similarly, that the width of the meson-meson resonance increases with $g$ is also a
consequence of the contraction of the mesons caused by the stronger quark-antiquark
attraction. This shrinking in size leads to a larger spreading of the meson-$q\bar q$ vertex
function in momentum space (see Fig.\ \ref{wavefunctions}), which in turn contributes to the
overlap integrals in the included Feynman diagrams in a wider momentum range, thereby
broadening the resonance.

Finally, we should mention that our principal finding -- the existence of a narrow low-lying
resonance in our calculations -- does not depend on the particular value chosen as a
constraint for the $q \bar q$ bound-state mass. If we use, instead of 140 MeV, a much larger
or a much smaller value, we find again a narrow resonance close to threshold.

\section{Conclusions}
\label{S:conclusions}

We calculated various models for quark-antiquark vertex  functions within the 't
Hooft model by solving the corresponding $q \bar{q}$ Bethe-Salpeter equation. In all
cases, the bare quark masses and the quark-gluon coupling constants  were tuned such
that the mass of the $q \bar{q}$ bound state  coincides with the pion mass.

We found that, within a limited range of coupling constants and 
bare quark masses, one obtains $q \bar{q}$ bound states with the
features of a constituent quark model, i.\ e., where the meson mass is
approximately equal to the sum of the dressed quark masses.
On the other hand, for larger values of the quark-gluon coupling constant this
constituent quark picture is no longer sustained.

We used the calculated Bethe-Salpeter wave functions to derive meson-meson
(``pion-pion'') scattering amplitudes within the 't Hooft model. They are
calculated  from the leading-order quark-exchange diagrams. These QCD-based
meson-meson amplitudes exhibit a resonance structure close to threshold in all
considered cases. We extracted an effective mass and width of this ``sigma-like''
resonance by comparison with a simple  s-channel resonance model where ``pions'' and
``sigmas'' instead of quarks are  the effective degrees of freedom.

The extracted values are not meant to be completely  realistic in the sense that
they should reproduce the experimental data, since the 't Hooft model is QCD only
under simplifying assumptions. However, the results of this calculation demonstrate
that the 't Hooft model can accomodate a resonance for the four-quark system,
already in the leading-order quark-exchange processes. In those exchange mechanisms,
the diquark correlation, given by the quark-antiquark vertex function, plays an
important role in determining the energy dependence of the meson-meson scattering
amplitude. 

The fact that in our 't Hooft model calculations a narrow resonance lies close to
threshold, while the broader sigma resonance is supposed to be located at higher
energy, is indicative to the limitations of the 't Hooft model. It also hints at the
importance of higher-order quark exchange processes as well as of gluon exchange
contributions (higher order terms in the $1/N_c$ expansion), which should be
investigated in future work.

\section{Acknowledgements}

This work was  supported by the Funda\c c\~ao para 
a Ci\^encia e Tecnologia, Portugal, and
FEDER, under grant number SFRH/BPD/5661/2001 (Z.B.), CERN/FIS/43709/2001 and
POCTI/FNU/40834/2001 (M.T.P and A.S.).

\end{document}